\documentclass[prl,aps,twocolumn,amsmath,amssymb]{revtex4}

\usepackage{graphicx}
\usepackage{dcolumn}
\usepackage{bm}

\begin{document}

\preprint{APS/123-QED}

\title{Coherent probing of excited quantum dot states in an interferometer}

\author{Martin Sigrist,$^1$ Thomas Ihn,$^1$ Klaus Ensslin,$^1$ Matthias Reinwald,$^2$ and Werner Wegscheider$^2$}
\affiliation{
$^1$Solid State Physics Laboratory, ETH Z\"urich, 8093 Z\"urich, Switzerland\\
$^2$Institut f\"ur experimentelle und angewandte Physik,
Universit\"at Regensburg, Germany }

\date{\today}

\begin{abstract}
Measurements of elastic and inelastic cotunneling currents are presented on a two-terminal Aharonov--Bohm interferometer with a Coulomb blockaded quantum dot embedded in each arm. Coherent current contributions, even in magnetic field, are found in the nonlinear regime of inelastic cotunneling at finite bias voltage. The phase of the Aharonov--Bohm oscillations in the current exhibits phase jumps of $\pi$ at the onsets of inelastic processes. We suggest that additional coherent elastic processes occur via the excited state. Our measurement technique allows the detection of such processes on a background of other inelastic current contributions and contains information about the excited state occupation probability and the inelastic relaxation rates.

\end{abstract}

\maketitle
Quantum dots (QDs) in the Coulomb blockade regime show well understood conductance resonances at low bias voltage when the gate voltage is swept \cite{Kouwenhoven97}. In interference experiments involving the Aharonov--Bohm (AB) effect a coherent current contribution was observed on such resonances \cite{Yacoby1995}. At increased tunnel coupling, higher order cotunneling \cite{Nazarov92} leads to a finite conductance between these resonances. At low bias, elastic cotunneling occurs which is energy conserving. Elastic cotunneling has also been shown to have a coherent contribution \cite{Sigrist2006}.
At finite bias voltages, inelastic cotunneling is observed \cite{Franceschi2001} in which the tunneling process excites the QD. Inelastic cotunneling was used for studying Zeeman-splitting \cite{Kogan04} and the singlet--triplet gap \cite{Zumbuhl04}.

If the QD is embedded in an AB interferometer, inelastic processes are not expected to contribute to interference, because the resulting excited dot state allows which-path detection.
Here we present the observation of coherent contributions to the current at and beyond the onset of inelastic cotunneling. We show that the corresponding AB oscillations  exhibit a phase change of $\pi$ at the bias voltage of the inelastic onset in most cases. An explanation of this finding requires the contribution of additional coherent elastic cotunneling processes through the involved excited state. The experiments therefore demonstrate the possibility of probing excited states and elastic cotunneling processes through them via the coherent current contribution in the nonlinear bias regime.

\begin{figure}[b]
\centering
\includegraphics[width=3.1in]{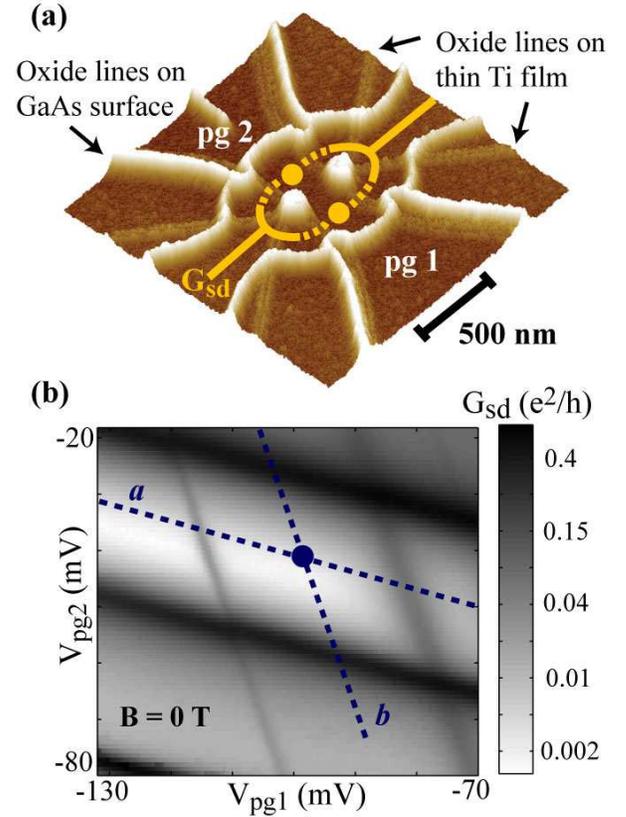}
\caption{\label{figure1} (a) SFM-micrograph of the structure (Details in the main text).
(b) $G_\mathrm{sd}$ as a function of $V_\mathrm{pg1}$ and $V_\mathrm{pg2}$ representing the
charge stability diagram of the two QDs.
The finite-bias measurements in Fig.\,\ref{figure2} were taken along
the dashed lines.}
\end{figure}

The sample shown in Fig.\,\ref{figure1}(a) is based on a Ga[Al]As
heterostructure with a two-dimensional electron gas (2DEG) 34~nm
below the surface. It was fabricated by multiple-layer local
oxidation with a scanning force microscope~\cite{Sigrist2004}: The
2DEG is depleted below the oxide lines written on the GaAs surface
[bright lines in Fig.\,\ref{figure1}(a)] thus defining the ring-interferometer. A Ti film evaporated on top is
cut by local oxidation [faint lines in Fig.\,\ref{figure1}(a)] into
mutually isolated top gates.

A QD is embedded in each arm of the resulting
AB-interferometer as indicated by the dots
in Fig.\,\ref{figure1}(a). Direct tunneling between the two
dots is suppressed by
applying a negative voltage between the 2DEG and the metallic
top gate, in contrast to previous experiments \cite{Sigrist2006}. In-plane gates pg1 and pg2 are used as plunger
gates for dot~1 and 2, respectively. Topologically
the sample is similar to those of Refs.~\onlinecite{Holleitner2001} and \onlinecite{Hatano2004}. More details about the sample are found
in Ref.~\onlinecite{Sigrist2006}. The source--drain two-terminal differential conductance, $G_{sd}=\partial I/\partial V_\mathrm{sd}$, was measured as indicated in Fig.\,\ref{figure1}(a) with low-frequency lock-in techniques at 120~mK
electronic temperature. 

With the dots strongly coupled to the
ring (open regime) and applying a magnetic field, $B$, normal to the 2DEG plane, we observe a periodically modulated conductance with an AB period of 22~mT, consistent with one magnetic flux quantum $\phi_0=h/e$ penetrating the area enclosed by the paths indicated in Fig.\,\ref{figure1}(a).

The conductance $G_\mathrm{sd}$ of the system in the Coulomb blockade regime of the dots is plotted as a function of $V_\mathrm{pg1}$ and $V_\mathrm{pg2}$ in
Fig.\,\ref{figure1}(b). The two families of parallel dark lines
differing in slope are conductance resonances of dot~1 and dot~2.
There is no apparent avoided crossing between resonances due to the absence of tunnel coupling and an interdot/intradot capacitance ratio of less than 1/20. From the resonance heights we estimate that the coupling of dot~2 to the leads is
stronger by more than one order of magnitude than that of dot~1. This regime differs completely from the experiments in Ref.~\onlinecite{Sigrist2006}, because direct tunneling between the dots is absent and their coupling to the ring is much stronger. 

\begin{figure}[t]
\centering
\includegraphics[width=3.1in]{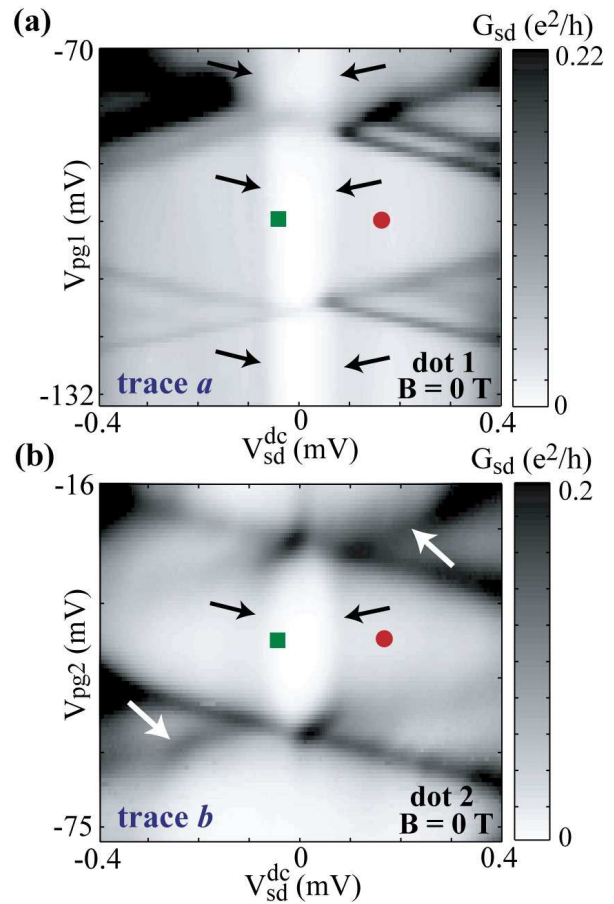}
\caption{\label{figure2} (a) Differential conductance is measured
as a function of DC source-drain voltage along trace \emph{a} in
Fig.\,\ref{figure1}(b). An inelastic onset independent of the
electron number of dot~1 is superposed on the Coulomb diamonds.
(b) Differential conductance is measured as a function of bias
voltage along trace \emph{b} in Fig.\,\ref{figure1}(b). The
inelastic onset can be linked to an excited state of dot~2.}
\end{figure}

Along the dashed lines `a' and `b' in Fig.\,\ref{figure1}(b) we measured
$G_\mathrm{sd}(V_\mathrm{sd})$. Along line `a' the electron number changes
in dot~1 while it is constant in dot~2, an vice versa along line `b'. The corresponding Coulomb blockade diamonds shown in Fig.\,\ref{figure2} give a charging energy of about 0.7~meV and single-particle level spacings of about 0.1~meV.

The cotunneling current observed at the intersection of lines `a' and `b' in Fig.\,\ref{figure1}(b) can be seen in Fig.\,\ref{figure2}. It shows $V_\mathrm{sd}$ thresholds for inelastic cotunneling in one of the two QDs.
In Fig.\,\ref{figure2}(a) we observe a superposition of
Coulomb diamonds and an inelastic cotunneling onset at about
$|V_\mathrm{sd}^{dc}|=0.1$~meV (black arrows). It
persists when the electron number in dot~1 is changed. The same onset, but now observed along trace `b', is seen only in the central diamond, i.e. it depends on the electron number in dot~2 [Fig.\,\ref{figure2}(b)].
We conclude that inelastic cotunneling occurs in dot~2 beyond
the bias threshold.

The inelastic cotunneling onset connects to excited state
resonances outside the Coulomb-blockaded region [white arrows in Fig.\,\ref{figure2}(b)]. However, only resonances with positive slope are observed and the corresponding resonances with negative slopes are missing,
indicating asymmetric tunnel coupling. We have therefore fine-tuned
the tunnel barriers in order to reach a $G_\mathrm{sd}$ trace as symmetric as possible in $V_\mathrm{sd}$ [Fig.\,\ref{figure4}(a)].

\begin{figure}[t]
\centering
\includegraphics[width=3.1in]{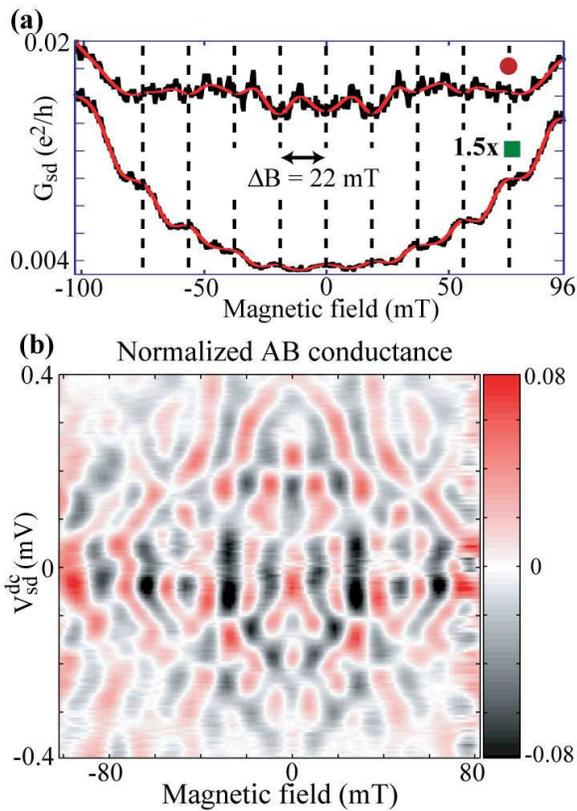}
\caption{(a) $B$ dependence of the differential
conductance for gate voltages set to the center of the hexagon in
Fig.\,\ref{figure1}(b). Bottom trace: low DC bias voltage (square, c.f. Fig.\,\ref{figure2}), top trace: high DC bias voltage (dot).
(b) Normalized AB conductance $g_\mathrm{AB}(B,V_\mathrm{bias})$.
}
\label{figure3}
\end{figure}

Measurements of the AB effect in a magnetic field allow
the detection of phase-coherent contributions to the cotunneling current.
We have measured $G_\mathrm{sd}$ as a function of $B$ at the crossing point of lines `a' and `b' in Fig.\,\ref{figure1}(b) for a number of DC source--drain voltages. Two of these are displayed in Fig.\,\ref{figure3}(a). The lower trace corresponds to low DC bias voltage as marked by a square in
Fig.\,\ref{figure2}. AB oscillations with a maximum at zero $B$
and a period of 22~mT are observed confirming a phase-coherent contribution to the elastic cotunneling current.

The upper trace in Fig.\,\ref{figure3}(a) taken at higher
DC bias voltage [dot in Fig.\,\ref{figure2}] involves inelastic cotunneling through dot~2. Also in this case AB oscillations are observed, but show a minimum at $B=0$.
We find either maxima or minima at $B=0$, i.e., phase rigidity, for all investigated source--drain voltages (see below), in contrast to non Coulomb blockaded systems \cite{Leturcq06}.
It is evident from the data that the participation of the inelastic cotunneling process does not hamper the occurrence of quantum interference. We emphasize that $G_\mathrm{sd}$ does not detect the total (energy integrated) DC current, but only a small (compared to temperature) energy window around the chemical potentials in source and drain.

We analyze the data following Ref.\,\onlinecite{Sigrist2004a} by splitting the measured $G_\mathrm{sd}(B)$ into three additive contributions: a smoothly varying background conductance $G_\mathrm{bg}(B,V_\mathrm{sd})$, the coherent AB-contribution $G_\mathrm{AB}(B,V_\mathrm{sd})$, and a contribution with fluctuations much faster than the AB period. In Fig.\,\ref{figure3}(a) we have plotted $G_\mathrm{bg}+G_\mathrm{AB}$ (smooth, gray) on top of the measured $G_\mathrm{sd}$ traces (ragged, black).
Small conductance fluctuations beyond the AB frequency that may arise due to interference effects in the contacts outside the system are filtered out with this procedure.

\begin{figure}[t]
\centering
\includegraphics[width=3.1in]{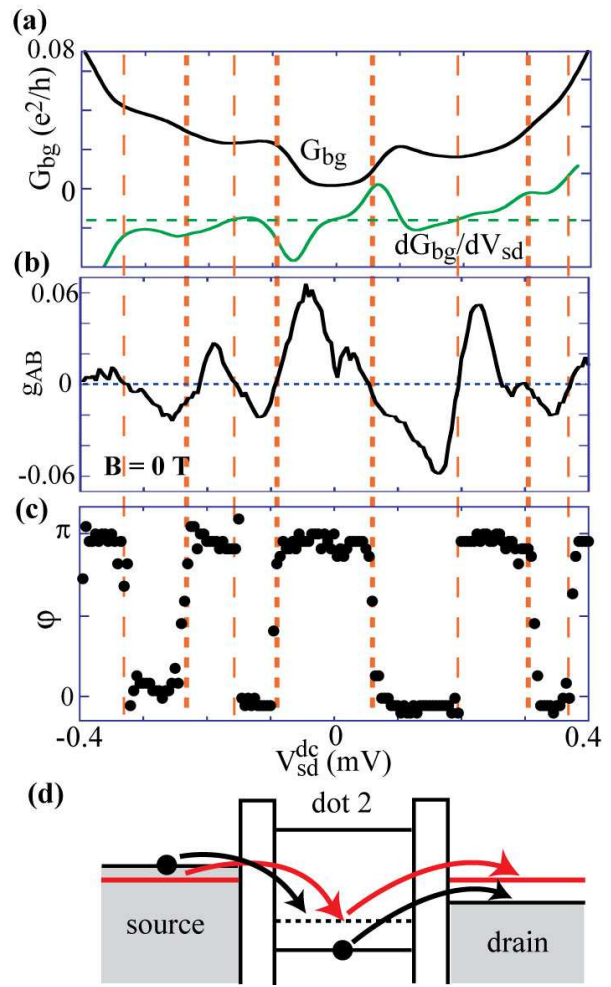}
\caption{\label{figure4} (a) Differential conductance and its derivative averaged over one AB period around zero magnetic field as a function of DC source-drain voltage.
(b) Normalized amplitude of the AB oscillation at $B=0$ as a function
of source--drain bias.
(c) AB-phase at $B=0$.
(d) Schematic of elastic cotunneling transport through dot
2 triggered at the inelastic onset.}
\end{figure}
Figure~\ref{figure3}(b) displays the normalized AB conductance $g_\mathrm{AB}(B,V_\mathrm{sd})=G_\mathrm{AB}(B,V_\mathrm{sd})/G_\mathrm{bg}(B,V_\mathrm{sd})-1$. This quantity can take values in the interval $[-1,1]$ and, evaluated at an AB-maximum or minimum, its modulus is related to the visibility of the AB oscillations. The visibility found in the measurement is always less than 0.1, a value comparable to other experiments \cite{Sigrist2004a,Yacoby1995}, but significantly lower than that observed in Ref.~\onlinecite{Sigrist2006} where the tunnel coupling between the QDs was significant.

At zero magnetic field, $g_\mathrm{AB}$ in Fig.\,\ref{figure3}(b) shows either maxima or minima [see also Fig.\,\ref{figure3}(a)].
Fig.\,\ref{figure4}(c) shows $\varphi(B=0)$ of the oscillations as determined from a fit of $a\cos(\omega_\mathrm{AB}B+\varphi)$ to the data around $B=0$, with amplitude $a$ and phase $\varphi$ being fitting parameters, and $\omega_\mathrm{AB}=2\pi BA/\phi_0$ ($A$ is the ring area). Several phase
jumps between the two values $0$ and $\pi$ are observed which correspondingly appear in Fig.\,\ref{figure3}(b) with changing $V_\mathrm{sd}$.
The generalized Onsager symmetries imposed on the two-terminal measurement
restrict the AB phase only at $B=0$ and at low bias to be
either zero or $\pi$ \cite{Leturcq06}. The measurement shows that $\varphi$ in our system is at $B=0$ very close to 0 or $\pi$ even in the nonlinear regime. 

Figures~\ref{figure4}(a)--(c) relate $G_\mathrm{bg}(B=0,V_\mathrm{sd})$, $g_\mathrm{AB}(B=0, V_\mathrm{sd})$ and $\varphi(B=0,V_\mathrm{sd})$. $G_\mathrm{bg}$ in (a) shows an elastic cotunneling contribution at low bias and an inelastic cotunneling onset slightly below $\left|V_\mathrm{sd}\right|=0.1$~V. Additional weaker shoulders in $G_\mathrm{bg}$ can be well detected as extrema in the derivative of $G_\mathrm{bg}$ shown in the same plot.

Whenever $g_\mathrm{AB}(B=0,V_\mathrm{sd})$ in Fig.\,\ref{figure4}(b) crosses zero, the phase in Fig.\,\ref{figure4}(c) jumps rather abruptly between $0$ and $\pi$. The dashed vertical lines in Figs.~\ref{figure4}(a)--(c) indicate that a correlation exists between some of these phase jumps and the inelastic cotunneling shoulders in $G_\mathrm{bg}$.
At small $V_\mathrm{sd}$ there is
an AB maximum ($\varphi=\pi$), but the phase jumps to $0$ at the first inelastic onset for both polarities. The AB phase jumps back to $\pi$ (AB maximum) at further increased $|V_\mathrm{sd}|$. Another phase jump is observed at the second inelastic onset at about
$|V_\mathrm{sd}|=0.3$~meV. Again, $\varphi$ jumps back by
increasing $|V_\mathrm{sd}|$ further.
Summarizing, we find the same AB phase for each of the two
inelastic onsets with different AB phases in-between. From measurements of the same sample in different regimes we can say that most inelastic cotunneling onsets lead to a $\pi$ phase jump in the AB oscillations of the differential conductance, although there are occasional exceptions where no phase jump can be observed.


The experiment raises the question why quantum coherence is not impaired by the presence of inelastic cotunneling. Leaving dot 2 in an excited state after such a cotunneling event allows which-path detection. A possible scenario resolving this puzzle is shown in Fig.\,\ref{figure4}(d). An inelastic cotunneling process excites the dot and increases the occupation probability of the excited state. Starting from this state, coherent {\em elastic} cotunneling processes via the {\em excited} state in dot 2 can take place that interfere with elastic cotunneling processes through dot~1 and give rise to the observed AB oscillations.

For such processes to occur, a significant population of the excited state is required. The relaxation rate {\em from} the excited state to the ground state (by phonon emission or further inelastic electron tunneling) must be small compared to the rate bringing the QD from the ground to the excited state via inelastic cotunneling. Charge relaxation times in QDs have been measured to be of the order of $1-10$~ns and attributed to acoustic phonon emission \cite{Fujisawa02}. Relaxation times involving a spin-flip can be much longer \cite{Fujisawa02,Hanson05}. Inelastic cotunneling relaxing the dot back to the ground state will have a similar time scale as the process exciting the dot.

Once the above condition is fulfilled, elastic cotunneling through the excited state can take place. We estimate its contribution to the differential conductance to be typically comparable to that of zero bias elastic cotunneling through the ground state
and to the inelastic contributions.
This discussion makes clear that the coherent contribution to the tunneling current probes the occupation probability of the excited QD state and thereby gives information about the rates of inelastic processes.
The scenario proposed here is the cotunneling analogue to the cotunneling mediated transport through excited states in the Coulomb-blockade regime reported recently \cite{Schleser2005}. It can be particularly strong, if the excited state transition has a significantly stronger tunnel coupling to the leads than the ground state transition. This is supported by the fact that we did not find AB oscillations in the regime of weak interdot {\em and} weak dot--ring coupling.

Our experiment differs significantly from previous measurements addressing the electrostatic AB effect \cite{Nazarov1993}. A recent experiment on an AB ring \cite{vanderWiel2003} was interpreted in terms of this prediction, and an experiment on a Mach-Zender interferometer obtained similar results \cite{Neder2006}. Phenomenologically, these results show similar abrupt jumps by $\pi$ in the AB phase and oscillations of the visibility with $V_\mathrm{sd}$. An important property of our structure is the presence of the two quantum dots with discrete levels, which allows only cotunneling currents to flow. The close relation of some phase jumps to the addition of transport channels through one of the two dots is unlikely to occur by chance as a result of the electrostatic AB effect.

In conclusion, we have shown that the measurement of the coherent contribution to the cotunneling current in an Aharonov--Bohm interference experiment can be used to detect coherent elastic cotunneling processes on a background of other inelastic processes. This coherent current contribution contains information about the occupation probability of the involved excited dot state and relaxation times. The results give a new perspective on inelastic cotunneling onsets. The measurement technique can be employed for further studies of coherent tunneling and interference involving quantum dots.

\begin{acknowledgments}
We thank Y. Meir for valuable discussions and appreciate financial support from
the Swiss National Science Foundation (Schweizerischer
Nationalfonds).
\end{acknowledgments}


\begin{thebibliography}{100}

\bibitem{Kouwenhoven97} L.P. Kouwenhoven {\it et al.},
in {\it Mesoscopic Electron Transport}, edited by . L.P. Kouwenhoven, G. Sch\"on, and L.L. Sohn, NATO ASI, Ser. E, Vol. 345 (Kluwer, Dordrecht, 1997), pp. 105--214.

\bibitem{Yacoby1995} A. Yacoby, M. Heiblum, D. Mahalu, H. Shtrikman, Phys. Rev. Lett. {\bf 74}, 4047 (1995).

\bibitem{Nazarov92} D.V. Averin, Yu.V. Nazarov, in {\it Single Charge Tunneling: Coulomb Blockade Phenomena in Nanostructures}, edited by H. Grabert, and M.H. Devoret (Plenum Press and NATO Scientific Affairs Division, New York, 1992), p. 217.

\bibitem{Sigrist2006}
M. Sigrist {\em et al.}, Phys. Rev. Lett. {\bf 96}, 036804 (2006).


\bibitem{Franceschi2001}
S. De Franceschi {\em et al.}, Phys. Rev. Lett. {\bf 86}, 878
(2001).

\bibitem{Kogan04} A. Kogan {\it et al.}, Phys. Rev. Lett. {\bf 93}, 166602 (2004).

\bibitem{Zumbuhl04} D.M. Zumbuhl {\it et al.}, Phys. Rev. Lett. {\bf 93}, 256801 (2004).

\bibitem{Sigrist2004}
M. Sigrist {\em et al.}, Appl. Phys. Lett. {\bf 85}, 3558 (2004).

\bibitem{Holleitner2001}
A.W. Holleitner {\it et al.},
Phys. Rev. Lett. {\bf 87}, 256802 (2001).

\bibitem{Hatano2004}
T. Hatano {\em et al.}, Phys. Rev. Lett. {\bf 93}, 066806 (2004);
M.C. Rogge {\em et al.}, Appl. Phys. Lett. {\bf 83}, 1163 (2003).



\bibitem{Leturcq06} R. Leturcq, D. Sanchez, G. G\"otz, T. Ihn, K. Ensslin, D.C. Driscoll, A.C. Gossard, Phys. Rev. Lett. {\bf 96}, 126801 (2006).

\bibitem{Sigrist2004a}
M. Sigrist {\em et al.}, Phys. Rev. Lett. {\bf 93}, 66802 (2004).

\bibitem{Fujisawa02} T. Fujisawa, D.G. Austing, Y. Tokura, Y. Hirayama, S. Tarucha, Nature {\bf 419}, 278 (2002).

\bibitem{Hanson05} R. Hanson {\it et al.}, Phys. Rev. Lett. {\bf 94}, 196802 (2005).

\bibitem{Schleser2005}
R. Schleser, T. Ihn, E. Ruh, K. Ensslin, M. Tews, D. Pfannkuche, D.C. Driscoll, A.C. Gossard, Phys. Rev. Lett. {\bf 94}, 206805 (2005).

\bibitem{Nazarov1993}
Y.V. Nazarov, Phys. Rev. B {\bf 47}, 2768 (1993).

\bibitem{vanderWiel2003}
W.G. van der Wiel, Y.V. Nazarov, S. de Franceschi, T. Fujisawa, J. Elzerman, E.W.G.M. Huizeling, S. Tarucha, L.P. Kouwenhoven, Phys. Rev. B {\bf 67}, 033307 (2003).

\bibitem{Neder2006}
I. Neder, M. Heiblum, Y. Levinson, D. Mahalu, V. Umansky, Phys. Rev. Lett. {\bf 96}, 016804 (2006).
\end{thebibliography}
\end{document}